\documentclass[pra,11pt,onecolumn, showkeys,showpacs]{revtex4}

\usepackage{times}

\begin{document}

\title{Cryptanalysis and improvement of dynamic quantum secret sharing protocol based on two-particle transform of Bell states}
\author{Gan Gao$^{1,2}$ \\
\normalsize $^{1}$ Department of Electrical Engineering, Tongling
University, Tongling 244000, China\\
\normalsize $^{2}$ Engineering Technology Research Center of
Optoelectronic Technology Appliance, \\
\normalsize Tongling University, Tongling 244000, China\\}
%\affiliation{a}
\date{\today}
\maketitle

\begin{minipage}{420pt}

{\bf Abstract}

In this paper [Chin. Phys. B {\bf 27} (2018) 080304], Du and Bao proposed a quantum secret sharing protocol based on two-particle transform of Bell states.
We study the security of the proposed protocol and find that it is not secure, that is, the two dishonest agents, Bob and Zach, can collude to obtain Alice's secret messages without
the help of the other agents. Finally, we give a possible improvement of the
proposed protocol.\\
\



\end{minipage}\\
\noindent {\bf 1. Introduction }\\

The cryptography is playing a significant role in the information society,
and can be classified into the classical cryptography and the quantum cryptography.
The main difference between the two cryptography focuses on the protection of security.
The security of classical cryptography depends on the
computational complexity. However, this kind of computational complexity might be broken by the strong power of advanced algorithms.
The security of quantum cryptography depends on the principles of quantum mechanics,
and can guarantee the unconditional security not only theoretically but also in a actual implementation.
Thus far, many branches of quantum cryptography have been presented to offer various security properties, including quantum
key distribution (QKD)[1-9], quantum secure direct communication (QSDC)[10-26], quantum secret sharing (QSS) [27-53], and so on.
In the following, let us introduce the three listed branches one by one.
QKD, which is the earliest and maturest branch, is a process in which
two communication parties first generate a shared secret key by
quantum states, and then apply this key to encrypt and decrypt the secret
messages. Since Bennett and Brassard [1] introduced the first QKD protocol
with nonorthogonal single polarization states, all kinds of QKD protocols were put forward. For example, Deng and
Long [2] proposed a two-step QKD protocol using practical faint laser pulses.
Boyer {\it et al.} [3] proposed a QKD scheme in which one participant owns the quantum device and the other doesn't.
Li {\it et al.} [4] proposed two QKD protocols over two different collective-noise channels.
Gao [5] proposed a QKD protocol by swapping the entanglement of $\chi$-type states.
Lo {\it et al.} [7] proposed a measurement-device-independent QKD protocol, and so on.
Different from QKD, QSDC is to directly transmit secret messages
without first generating a key to encrypt them. In 2002, Long
and Liu [10] proposed the first QSDC protocol using the
concept of quantum data block. In 2004, Deng and Long [11] proposed
a QSDC protocol using only a sequence of single photons.
In 2005, Wang {\it et al.} [12] proposed a
QSDC protocol with quantum superdense coding in high-dimension
Hilbert space. In 2007, Li {\it et
al.} [15] proposed a QSDC protocol with quantum encryption by using
pure entanglement states. In 2013, Ren {\it et
al.} [22] proposed a robust QSDC protocol with the spatial-mode
entanglement of two-photon systems, and so on.
In general, there exist only two communication parties in QKD and QSDC.
However, there are at least three communication parties in QSS.
In 1999, Hillery, Buz\v{e}k and Berthiaume [27] used
three-particle GHZ state and four-particle GHZ state to propose the first QSS protocol.
In 2003, Bagherinezhad and Karimipour [28] utilized reusable GHZ states as secure carriers
to propose a QSS protocol. In 2006,
Deng {\it et al.} [32] proposed a circular QSS protocol in which the quantum information carrier, single photons
or entangled particles, can circularly run.
In 2008, Markham {\it et al.} [36] gave a unified approach to secret sharing
of both quantum and classical secrets using graph states.
In 2012, Jia {\it et al.} [45] proposed two dynamic QSS protocols
in which the change of the agent group is allowable during the procedure of sharing information.
In 2017, Wang {\it et al.} [51] proposed a secure ($k,n$)-threshold QSS protocol
based on local distinguishability of orthogonal multiqudit entangled states, and so on.

Recently, Du and Bao [52] proposed a novel multiparty QSS protocol (hereafter called DB protocol).
It is interesting that the DB protocol uses the two-particle transform of Bell states
and has the functions of dynamic parameter update. However, it is somewhat a
pity that there exists a security loophole in the DB protocol.
That is, the two dishonest agents, Bob and Zach, can collaborate to obtain Alice's secret messages without being detected.\\

\noindent {\bf 2. Security loophole in the DB protocol}\\

In order to clearly show the security loophole, firstly, let us review the five-party case of the DB protocol [52] as follows.

(1) Alice prepares $k$ pairs of Bell states (i.e.,$|\varphi_{1}\rangle_{th},...,|\varphi_{k}\rangle_{th}$), where each pair is randomly in $\{|\phi^{\pm}\rangle=(|00\rangle\pm|11\rangle)/\sqrt{2},|\psi^{\pm}\rangle=(|01\rangle\pm|10\rangle)/\sqrt{2}\}$.
She takes out photons $t$ and $h$ of these Bell states to form T-sequence and H-sequence, respectively.
Then, the T-sequence is sent to Bob.

(2) After receiving the T-sequence, firstly, Bob checks whether it is composed of single photons.
Then, he performs the local unitary operation $U(\alpha_{i})=\cos\alpha_{i}|0\rangle\langle0|+\cos\alpha_{i}|1\rangle\langle1|-\sin\alpha_{i}|1\rangle\langle0|+\sin\alpha_{i}|0\rangle\langle1|$ on photon $t_{i}$. Here, $i=1,2,...,k$ and $\alpha_{i}\in \{0,\frac{2\pi}{3},\frac{4\pi}{3}\}$.
Lastly, Bob sends the T$^{(1)}$-sequence, which is transformed from the T-sequence, to Charlie.

(3) After Charlie receives the T$^{(1)}$-sequence, what he does is the same as what Bob does.
Then Charlie sends the T$^{(2)}$-sequence, which is transformed from the T$^{(1)}$-sequence, to Green.
What Green does is also the same as what Bob does. Then he sends the T$^{(3)}$-sequence,
which is transformed from the T$^{(2)}$-sequence, to Zach. Zach also performs the local unitary
operation $U(\alpha_{i})$ on photon $t_{i}$ of his receiving sequence and remains the T$^{(4)}$-sequence
which is transformed from the T$^{(3)}$-sequence.

(4) Alice performs the four Pauli operations
($\sigma_{00}=|0\rangle\langle0|+|1\rangle\langle1|$,
$\sigma_{01}=|0\rangle\langle1|+|1\rangle\langle0|$,
$\sigma_{10}=|0\rangle\langle0|-|1\rangle\langle1|$,
$\sigma_{11}=|0\rangle\langle1|-|1\rangle\langle0|$) on photons of the H-sequence to encode her secret M.
Then she sends the H$^{(1)}$-sequence, which is transformed from the H-sequence, to Zach.

(5) After Zach receives the H$^{(1)}$-sequence, all the agents and Alice start to check eavesdropping.
First, Alice randomly selects $k_{1}$ positions of the T$^{(4)}$-sequence and tells the selected positions to all the agents.
Then, all the agents choose Green to collect the others' messages of operations on $k_{1}$ positions and
to perform the reverse compound operations. Next, Green performs
Bell-basis measurements on two corresponding photons in $k_{1}$ positions of both the T$^{(4)}$-sequence and the H$^{(1)}$-sequence,
and tells his measurement outcomes to Alice. In result, Alice can judge whether
the eavesdropping exists or not. If no eavesdropping exists, Alice will announce all of the initial Bell states.
So all the agents can collaborate to recover Alice's secret M.

We can see that, in the DB protocol, the local unitary operation performed by each agent is chosen from
the phase shift operation set $S=\{U(0),U(2\pi/3),U(4\pi/3)\}$. Since the three operations cannot
be exactly distinguished by measuring the different quantum states, Du and Bao stated that their QSS protocol
was secure. However, this is not a fact. In what follows, we will prove that the above five-party case is not secure by
designing a attack strategy on it. Our attack strategy, which is implemented by the two dishonest agents, Bob and Zach, is described as follows.

In advance, Bob and Zach prepare some Bell states, where each is $|\psi^{+}\rangle_{t'h'}=\frac{1}{\sqrt{2}}(|01\rangle_{t'h'}+|10\rangle_{t'h'})$.
According to the forming-sequence manner in the above step (1), they also get the two sequences, T$'$-sequence and H$'$-sequence. Here,
the T$'$-sequence is in Bob's hand and the H$'$-sequence is in Zach's hand. In the above step (2), after Bob receives the T-sequence from Alice,
he doesn't perform any operations on it, but secretly sends it to Zach. In addition, he performs the local unitary operation $U(\alpha_{i})$
on photon $t'_{i}$ of the T$'$-sequence, and sends the T$'^{(1)}$-sequence, which is transformed from the T$'$-sequence, to Charlie. At this moment,
the T$^{(1)}$-sequence has been replaced with the T$'^{(1)}$-sequence, which is not known by Charlie, Green and Alice.
After receiving the T$'^{(1)}$-sequence, Charlie performs the local unitary operation $U(\alpha_{i})$ on it {\it as of old}, and sends the T$'^{(2)}$-sequence,
which is transformed from the T$'^{(1)}$-sequence, to Green. After Green receives the T$'^{(2)}$-sequence, what he needs to do is the same as what Charlie does.
This means that Zach will receive the T$'^{(3)}$-sequence, which is transformed from the T$'^{(2)}$-sequence, from Green. After receiving the sequence, he also performs the local unitary operation
$U(\alpha_{i})$ on it. Now, Zach holds the three sequences: the T$'^{(4)}$-sequence (transformed from the T$'^{(3)}$-sequence),
the H$'$-sequence and the T-sequence. As soon as Alice sends the H$^{(1)}$-sequence to him, he will hold all the sequences.
When Alice announces $k_{1}$ positions of the T$^{(4)}$-sequence, Zach immediately performs Bell-basis measurements on two corresponding photons
in $k_{1}$ positions of both the T-sequence and the H$'$-sequence. Obviously, there exists a process of swapping entanglement. Let us give an example to show this process. Suppose that Alice's unitary operation on photon $h_{k^{j}_{1}}$ (the subscript $k^{j}_{1}$ denotes the $j$th in $k_{1}$ positions) and her initial Bell state are $\sigma_{01}$ and $|\psi^{-}\rangle_{t_{k^{j}_{1}}h_{k^{j}_{1}}}$, respectively, and Bob's, Charlie's, Green's and Zach's local unitary operations on photon $t'_{k^{j}_{1}}$ are $U(2\pi/3)$, $U(0)$, $U(2\pi/3)$ and $U(4\pi/3)$, respectively. When Zach performs Bell-basis measurement on photons $t_{k^{j}_{1}}$ and $h'_{k^{j}_{1}}$, the system evolves as follows:

\begin{displaymath}
(\sigma_{01}|\psi^{-}\rangle_{t_{k^{j}_{1}}h_{k^{j}_{1}}})\otimes (U(4\pi/3)U(2\pi/3)U(0)U(2\pi/3)|\psi^{+}\rangle_{t'_{k^{j}_{1}}h'_{k^{j}_{1}}})=\frac{1}{2}(|\phi^{+}\rangle_{t_{k^{j}_{1}}h'_{k^{j}_{1}}}U(2\pi/3)|\psi^{-}\rangle_{t'_{k^{j}_{1}}h_{k^{j}_{1}}}+
\end{displaymath}
\begin{equation}
|\phi^{-}\rangle_{t_{k^{j}_{1}}h'_{k^{j}_{1}}}U(2\pi/3)|\psi^{+}\rangle_{t'_{k^{j}_{1}}h_{k^{j}_{1}}}-|\psi^{+}\rangle_{t_{k^{j}_{1}}h'_{k^{j}_{1}}}U(2\pi/3)|\phi^{-}\rangle_{t'_{k^{j}_{1}}h_{k^{j}_{1}}}-|\psi^{-}\rangle_{t_{k^{j}_{1}}h'_{k^{j}_{1}}}U(2\pi/3)|\phi^{+}\rangle_{t'_{k^{j}_{1}}h_{k^{j}_{1}}})
\end{equation}
According to equation (1), we see that Zach's Bell-basis measurement outcome is one of $|\phi^{+}\rangle_{t_{k^{j}_{1}}h'_{k^{j}_{1}}}$, $|\phi^{-}\rangle_{t_{k^{j}_{1}}h'_{k^{j}_{1}}}$, $|\psi^{+}\rangle_{t_{k^{j}_{1}}h'_{k^{j}_{1}}}$ and $|\psi^{-}\rangle_{t_{k^{j}_{1}}h'_{k^{j}_{1}}}$.
After the process of swapping entanglement is over, Zach makes a comparison for his Bell-basis measurement outcome and $|\psi^{-}\rangle_{t'h'}$ and obtains a unitary operation. This kind of Bell state comparison method and its comparison steps can be consulted in the papers [5,6].  Then, Zach performs the obtained unitary operation on photon $h_{k^{j}_{1}}$. At the same time, he performs Bell-basis measurements on two corresponding photons
in $k-k_{1}$ positions of both the T-sequence and the H$^{(1)}$-sequence. Since Green is chosen to collect messages, to perform the reverse compound operations and to perform Bell-basis measurements in the above step (5), this indirectly means that Zach needs to send two sequences to him. Notice that, in order not to be detected, the two sequences sent by Zach should be the T$'^{(4)}$-sequence and the H$^{(1)}$-sequence. Here, we can't help asking why Bob's and Zach's replacing action is not detected. In the following, we will give the reason by continuing to use the above example. Suppose that Zach's Bell-basis measurement outcome is $|\psi^{-}\rangle_{t_{k^{j}_{1}}h'_{k^{j}_{1}}}$, and he compares $|\psi^{-}\rangle_{t_{k^{j}_{1}}h'_{k^{j}_{1}}}$ with $|\psi^{+}\rangle_{t'h'}$ to obtain $\sigma_{10}$. According to equation (1), photons $t'_{k^{j}_{1}}$ and $h_{k^{j}_{1}}$ are in $U(2\pi/3)|\phi^{+}\rangle_{t'_{k^{j}_{1}}h_{k^{j}_{1}}}$. When $\sigma_{10}$ is performed on photon $h_{k^{j}_{1}}$, the system evolves as follows:
\begin{equation}
\sigma_{10}U(2\pi/3)|\phi^{+}\rangle_{t'_{k^{j}_{1}}h_{k^{j}_{1}}}=U(2\pi/3)|\phi^{-}\rangle_{t'_{k^{j}_{1}}h_{k^{j}_{1}}}
\end{equation}
Now, let us see which state photons $t_{k^{j}_{1}}$ and $h_{k^{j}_{1}}$ are in if Bob and Zach don't perform the replacing action.
When Bob's $U(2\pi/3)$, Charlie's $U(0)$, Green's $U(2\pi/3)$ and Zach's $U(4\pi/3)$ are performed on photon $t_{k^{j}_{1}}$ and Alice's $\sigma_{01}$ are performed on photon $h_{k^{j}_{1}}$, the system evolves as follows:
\begin{equation}
\sigma_{01}U(4\pi/3)U(2\pi/3)U(0)U(2\pi/3)|\psi^{-}\rangle_{t_{k^{j}_{1}}h_{k^{j}_{1}}}=U(2\pi/3)|\phi^{-}\rangle_{t_{k^{j}_{1}}h_{k^{j}_{1}}}
\end{equation}
That is, photons $t_{k^{j}_{1}}$ and $h_{k^{j}_{1}}$ are in $U(2\pi/3)|\phi^{-}\rangle_{t_{k^{j}_{1}}h_{k^{j}_{1}}}$. Obviously, the state of photons $t_{k^{j}_{1}}$ and $h_{k^{j}_{1}}$ is the same as that of photons $t'_{k^{j}_{1}}$ and $h_{k^{j}_{1}}$. Therefore, Bob's and Zach's replacing action cannot be detected, so that Alice thinks that the whole quantum channel is secure. Next, she announces all of the initial Bell states. As soon as Zach knows Alice's initial Bell states, plus the states that two corresponding photons
in $k-k_{1}$ positions of both the T-sequence and the H$^{(1)}$-sequence are in, he easily infers Alice's unitary operation, that is, her secret M.

In order to resist the above attack that Bob and Zach implement, we will give an improvement of the DB protocol. Here, this improvement begins with the fourth step because steps (1$'$), (2$'$) and (3$'$) in it are same as the former three steps in the DB protocol. (4$'$) Alice randomly selects some photons from the H-sequence and randomly uses the basis $\{|0\rangle,|1\rangle\}$ or $\{|+\rangle=\frac{|0\rangle+|1\rangle}{\sqrt{2}},|-\rangle=\frac{|0\rangle-|1\rangle}{\sqrt{2}}\}$ to measure each selected photon. Then, Alice announces the positions of the selected photons in the H-sequence, and asks Zach to send the partner photons of the selected photons in the T$^{(4)}$-sequence to her and all the agents to publish their local unitary operations in a random order. Next, Alice performs the reverse compound operations on the partner photons, and then measures the partner photons with the same basis that are used when measuring the selected photons in the H-sequence. According to her measurement outcomes, Alice can judge whether the eavesdropping exists or not. If no eavesdropping exists, Alice performs the four Pauli operations on photons of the H-sequence to encode her secret M and sends the H$^{(1)}$-sequence (transformed from the H-sequence) to Zach. (5$'$) This step is the same as step (5) in the DB protocol.

We see that another process to check the security is added in the improvement. This process is mainly used to prevent two dishonest agents from eavesdropping, which had been shown in Wang {\it et al.}'s improving QSS protocol [34]. Of course, in the DB protocol, the attack from a dishonest agent is also discussed (please see Section 4.2.2 in the paper [52]), but Du and Bao only analyze a special inside attack implemented by one dishonest agent, which is called a single attack customarily. For the joint attack that is implemented by two dishonest agents, they don't discuss while analyzing the security. As we all know, the joint attack has stronger attack power than the single attack because more messages may be utilized while cheating, which is also the reason that the DB protocol is insecure.
\\

\noindent {\bf 3. Conclusion}\\

In conclusion, we successfully show that, in the five-party case of the DB protocol, Bob and Zach can collude to obtain Alice's secret M without the help of the other agents, moreover, Bob's and Zach's eavesdropping action doesn't introduce any error. In other words, by designing a joint attack, the DB protocol is successfully proved to be insecure by us. In addition, in order to resist the joint attack, we make a modification for the DB protocol, that is, we give an improvement of the DB protocol. To the end, it is worth emphasizing that the above attack strategy is proposed by combining Bell state comparison and entanglement swapping, which is similar to that in the paper [42,46]. In addition, another attack strategy to combine Bell state comparison and quantum teleportation can also be seen in the papers [38,40]. So we hope that the application of Bell state comparison can be noticed in the future research on QSS.
\\

\noindent {\bf Acknowledgements}\\

I thank my parents for their encouragements.\\

\noindent {\bf References}

\noindent[1] C. H. Bennett, G. Brassard, in a {\it Proceedings of
the IEEE International Conference on Computers, Systems and Signal
Processings, Bangalore, India} (IEEE, New York, 1984) pp. 175-179.

\noindent[2] F. G. Deng, G. L. Long, Phys. Rev. A {\bf 70} (2004)
012311.

\noindent[3] M. Boyer, D. Kenigsberg and T. Mor, Phys. Rev. Lett. {\bf 99}
(2007) 140501.

\noindent[4] X. H. Li, F. G. Deng, H. Y. Zhou, Phys. Rev. A {\bf
78} (2008) 022321.

\noindent[5] G. Gao, Opt. Commun. {\bf 281} (2008) 876.

\noindent[6] G. Gao, Commun. Theor. Phys. {\bf 51} (2009) 820.

\noindent[7] H. K. Lo, M. Curty and B. Qi, Phys. Rev. Lett. {\bf 108}
(2012) 130503.

\noindent[8] L. H. Gong, H. C. Song, C. S. He, Y. Liu and N. R. Zhou, Phys. Scr. {\bf 89} (2014) 035101.

\noindent[9] X. Yang {\it et al.}, Phys. Rev. A {\bf 93} (2016) 052303.

\noindent[10] G. L. Long, X. S. Liu, Phys. Rev. A {\bf 65} (2002) 032302.

\noindent[11] F. G. Deng and G. L. Long, Phys. Rev. A  {\bf 69}
(2004) 052319.

\noindent[12] C. Wang, F. G. Deng, Y. S. Li, X. S. Liu and G. L.
Long, Phys. Rev. A   {\bf 71} (2005) 044305.

\noindent[13] C. Wang, F. G. Deng, G. L. Long, Opt. Commun. {\bf
253} (2005) 15.

\noindent[14] M. Lucamarini, S. Mancini, Phys. Rev. Lett. {\bf 94}
(2005) 140501.

\noindent[15] X. H. Li, C. Y. Li, F. G. Deng, P. Zhou, Y. J. Liang,
H. Y. Zhou, Chin. Phys. {\bf 16} (2007) 2149.

\noindent[16] S. Lin, Q. Y. Wen, F. Gao, F. C. Zhu, Phys. Rev. A  {\bf 78}
(2008) 064304.

\noindent[17] C. Wang, L. Xiao, W. Y. Wang, G. Y. Zhang and G. L.
Long, J. Opt. Soc. Am. B. $26$ (2009) 2072.

\noindent[18] G. Gao, M. Fang, Y. Wang and D. J. Zang, Int. J.
Theor. Phys. {\bf 50} (2011) 3089.

\noindent[19] T. J. Wang, T. Li, F. F. Du, and F. G. Deng, Chin.
Phys. Lett. {\bf 28} (2011) 040305.

\noindent[20] L. Dong, X. M. Xiu, Y. J. Gao, Y. P. Ren and H. W.
Liu, Opt. Commun.   {\bf 284} (2011) 905.

\noindent[21] S. H. Kao and T. Hwang, Chin. Phys. B {\bf 22} (2013) 060308.

\noindent[22] B. C. Ren, H. R. Wei, M. Hua, T. Li, F. G. Deng, Eur.
Phys. J. D  {\bf 67} (2013) 30.

\noindent[23] L. H. Gong, Y. Liu, N. R. Zhou, Int.
J. Theor. Phys. {\bf 52} (2013) 3260.

\noindent[24] Y. Chang, C. X. Xu, S. B. Zhang, L. L. Yan, Chin. Sci.
Bulletin {\bf 58} (2013) 4571.

\noindent[25] T. Y. Ye, Quantum Inf. Process. {\bf 14} (2015) 1469.

\noindent[26] W. Zhang {\it et al.}, Phys. Rev. Lett. {\bf
118} (2017) 220501.

\noindent[27] M. Hillery, V. Buz\v{e}k and A. Berthiaume,  Phys. Rev. A
{\bf 59}  (1999) 1829.

\noindent[28] S. Bagherinezhad and V. Karimipour, Phys. Rev. A {\bf 67}
(2003) 044302.

\noindent[29] F. L. Yan and T. Gao, Phys. Rev. A {\bf 72} (2005) 012304.

\noindent[30] Z. J. Zhang, Y. Li and Z. X. Man,  Phys. Rev. A
{\bf 71} (2005) 044301.

\noindent[31] H. F. Wang, X. Ji and S. Zhang, Phys. Lett. A {\bf 358} (2006) 11.

\noindent[32] F. G. Deng, H. Y. Zhou, G. L. Long, J. Phys. A {\bf 39}
(2006) 14089.

\noindent[33] Z. Y. Xue, Y. M. Yi and Z. L. Cao, Chin. Phys. B {\bf 15}
(2006) 01421.

\noindent[34]  T. Y. Wang, Q. Y. Wen, F. Gao, S. Lin, F. C. Zhu,
Phys. Lett. A {\bf 373} (2008)65.

\noindent[35] Y. Guo, G. H. Zeng, Z. G. Chen, Chin. Phys. Lett. {\bf 24}
(2007) 863.

\noindent[36] D. Markham and B. Sanders, Phys. Rev. A {\bf 78} (2008)
042309.

\noindent[37] Y. G. Yang, Q. Y. Wen, Chin. Phys. B {\bf 17} (2008) 0419.

\noindent[38] G. Gao, Opt. Commun. {\bf 282} (2009) 4464.

\noindent[39] B. Gu, C. Q. Li, F. Xu, Y. L. Chen, Chin. Phys.
B {\bf 18} (2009) 04690.

\noindent[40] Z. C. Zhu and Y. Q. Zhang, Chin. Phys. Lett. {\bf 27}
(2010) 060303.

\noindent[41] Y. G. Yang, W. F. Cao, Q. Y. Wen, Chin. Phys.
B {\bf 19} (2010) 050306.

\noindent[42] G. Gao, Opt. Commun. {\bf 283} (2010) 2997.

\noindent[43] S. Yang, X. B. Chen, Y. X. Yang, Commun. Theor. Phys. {\bf 58,} (2012) 51.

\noindent[44] X. B. Chen, S. Yang, Y. Su, Y. X. Yang, Phys. Scr. {\bf 86} (2012) 055002.

\noindent[45] H. Y. Jia, Q. Y. Wen, F. Gao, S. J. Qin and F. Z. Guo, Phys. Lett. A {\bf 376} (2012)
1035.

\noindent[46] G. Gao, Quantum Inf. Process. {\bf 12} (2013) 55.

\noindent[47] X. B. Chen, X. X. Niu, X. J. Zhou, Y. X. Yang, Quantum Inf. Process. {\bf 12} (2013) 365.

\noindent[48] M. M. Wang, X. B. Chen, Y. X. Yang, Quantum Inf. Process. {\bf 12} (2013) 785.

\noindent[49] M. M. Wang, X. B. Chen, Y. X. Yang, Quantum Inf. Process. {\bf 13} (2014) 429.

\noindent[50] M. M. Wang, W. Wang, J. G. Chen, A. Farouk, Quantum Inf. Process. {\bf 14} (2015) 4211.

\noindent[51] J. Wang, L. Li, H. Peng and Y. Yang, Phys. Rev. A {\bf 95}
(2017) 022320.

\noindent[52] Y. T. Du and W. S. Bao, Chin. Phys. B {\bf 27} (2018) 080304.

\noindent[53] X. B. Chen, X. Tang, G. Xu, Z. Dou, Y. L. Chen, Y. X. Yang, Quantum Inf. Process. {\bf 17} (2018) 225.

\enddocument